\documentclass[12pt,a4paper,notitlepage]{article}

\usepackage{authblk}
\usepackage{hyperref}
\usepackage{graphics,amsmath}
\usepackage{graphicx}
\usepackage{color}
\usepackage{xcolor}

\begin{document}

\title{Position-dependent mass effects in the electronic transport of two-dimensional quantum systems: applications to nanotubes}
%\title{}
%
\author[$\dagger$]{Felipe Serafim}
\author[$\star$]{F. A. N. Santos}
\author[$\dagger$]{Jonas R. F. Lima}
\author[$\ddag$]{Cleverson Filgueiras}
\author[$\dagger$]{Fernando Moraes}
\affil[$\dagger$]{Departamento de F\'{\i}sica, Universidade Federal Rural de Pernambuco, 52171-900, Recife, PE, Brazil}
\affil[$\star$]{ Departamento de Matem\'atica,  Universidade Federal  de Pernambuco, 50670-901, Recife, PE, Brazil}   \affil[$\ddag$]{Departamento de F\'{\i}sica,
Universidade Federal de Lavras, Caixa Postal 3037,
37200-000, Lavras, Minas Gerais, Brazil  }
\maketitle
   
\begin{abstract}
In this work, we investigate the electronic transport properties of curved two-dimensional quantum systems with a position-dependent mass (PDM). We find the Schr\"odinger equation for a general surface following the da Costa approach, obtaining the geometrical potential for systems with PDM. We obtain expressions for the transmittance and reflectance for a general surface of revolution and, as a first application of the general results obtained here, we investigate the transport properties of deformed nanotubes whose variation of the effective mass with the radius  has been disconsidered in previous studies.  We find that the inclusion of the position-dependent mass,   can induce a significant correction in the transport properties of the system. This reveals that the transport properties of two-dimensional quantum systems are sensitive to the PDM and that, when modeling electronic transport in surfaces, this effects should be considered.
\end{abstract}
\section{Introduction}
\label{intro}

The investigation of curved two-dimensional quantum systems is an important and interesting branch of condensed matter physics.  This fact has support on the ability of building non-planar quasi-two-dimensional substrates in desired shapes \cite{Mendach,coneexperiment}.  Therefore, from the theoretical point of view,  tools from differential geometry play an important role in the study of these systems.  For instance, in reference \cite{pnjunctions}, it is argued that surface curvature could provide the creation of p-n junctions. This way, the geometric control of local electronic properties in carbon nanoribbons and bilayer graphene sheets can be realized. Reflectionless transmission of a quantum particle across the catenoid (two-dimensional wormhole geometry) is predicted in \cite{Dandoloff}. Geometry can also induce a reminiscent of the Hall effect \cite{Atanasov}. Related important investigations can also be found in the literature, for instance: the quantum Hall effect near conical singularities \cite{conePRL} and investigations taking into account the Pauli equation for a charged spin particle on a curved surface with  externally applied fields \cite{paulicurved,truncado}. Recent developments can also be found in \cite{Xun2014132,Spittel:15,SHIKAKHWA20162876,SHIKAKHWA20161985,Jahangiri2016407,Panahi201657,Wang2016,centripetal,Lian}.

One important ingredient present in the study of  charge carriers in low-dimensional semiconductors is the {\it effective-mass approach} \cite{burt}.  More recently,  the interest in the study of the dynamics of a quantum particle with a position-dependent effective mass has grown since semiconductors with varying composition can be 
realized \cite{comp}.  A number of systems where  variable mass plays an important role has been reported in the literature.  A  few examples are: diffusion of particles with variable mass  \cite{Butko};   surface states of a one-dimensional semi-infinite crystal   \cite{Ren} and  scattering states of a quantum system in the presence of a Heaviside position-dependent mass jump \cite{Mustafa}. Furthermore,  a wide class of position-dependent mass oscillators and the corresponding coherent states is presented in reference \cite{Cruz} and,  in \cite{Gadella}, it is addressed the analysis of the influence of the delta-function well in systems with mass jumps.  

Even though the position-dependent mass (PDM) has been widely considered, it has been neglected in various studies  of two-dimensional curved quantum systems. In particular, it is well-know that the effective mass of the charge carriers in nanotubes depends, for instance, on the radius of the nanotube \cite{5175416}. However, in  previous studies of deformed nanotubes, where the radius of the nanotube changes with the position, the effective mass was considered constant \cite{Marchi,santos}. As described in these references, the curvature introduced by the deformations of the tubes affects the dynamics of charge carriers via the geometric potential introduced by da Costa \cite{dacosta} in 1981.

In this work, we generalize da Costa's formalism \cite{dacosta} for the quantum dynamics of a particle constrained to move on a curved surface, to include a position-dependent   effective mass.  To this aim, we derive the Schr\"{o}dinger equation for a quantum particle with variable mass constrained to move on a generic curved surface. Using da Costa's  thin-layer procedure, which gives rise to a quantum geometric potential \cite{dacosta,Ferrari},  we find that this potential keeps its original form. Nevertheless,   the Schr\"{o}dinger equation has now two extra pieces containing the derivatives of the mass. This way, the combined effects of extrinsic  geometry and variable mass introduce novel effects in comparison to the usual da Costa approach with constant mass. To make  clear the importance of theses changes, we apply the formalism developed here to the study of charge transport in the same deformed nanotubes  studied in Ref.  \cite{santos} with a constant effective mass. To do so, we derive  a system of coupled differential and algebraic  equations for ballistic charged particles  moving on a generic surface of revolution, which are solved numerically for the same surfaces studied in Ref.  \cite{santos}, for the sake of comparison. From the obtained transmittance versus incident energy plots it becomes evident that taking into account the dependence on position of the mass has important consequences.  For instance, the transmittance peaks shift to higher energy values while energy gaps become broader. In sum, this work provides an approach to describe two-dimensional quantum particles with position-dependent mass, constrained to move on curved surfaces, which can be used in a wide range of applications, including the design of novel electronic devices, using curvature as a tool.

This manuscript is organized as follows. In Section 2, we derive the Schr\"odinger equation for a spinless particle with variable mass, constrained to move on a generic curved surface, and  compare with the case where the mass is constant. In Section 3, we derive expressions for the transmittance and reflectance in terms of the boundary conditions at both ends of a  surface of revolution, which forms the deformed section of the nanotube. In section 4, we study the transport properties of deformed nanotubes with a position-dependent mass. The paper is summarized and concluded in Section 5.

%%%%%%%%%%%%%%%%%%%%%%%%%%%%%%%%%%%%%%%%%%%%
\section{The da Costa approach with PDM}
\label{sec:1}
%%%%%%%%%%%%%%%%%%%%%%%%%%%%%%%%%%%%%%%%%%%%%

Let us consider a particle with an effective variable mass $m^* (q_1,q_2)$ permanently attached to a surface $S$ of parametric equations $\overrightarrow{r}=\overrightarrow{r}(q_1, q_2)$, where $q_1$ and $q_2$ are coordinates on the surface. When the effective mass of the particle depends on position, and therefore is an operator, the usual kinetic energy operator is no longer Hermitian. A number of modifications to this operator has been  proposed in order to turn it Hermitian \cite{jonas}. Among them, a possibility considered by three of us (JRFL, CF and FM) in \cite{jonas} is to take into account all permutations among the operators in the kinetic energy. To the best of our knowledge, none of those proposals have been subjected to experimental tests. So, to illustrate our point of view that variable mass adds up to curvature influencing the dynamics of quantum particles confined to surfaces, we use the Hamiltonian proposed in \cite{jonas}. That is,
\begin{equation}
H_{kin}=-\frac{\hbar^2}{6}\left[\frac{1}{m^*}\bigtriangleup + \bigtriangledown \frac{1}{m^*} \bigtriangledown + \bigtriangleup \frac{1}{m^*} \right]  \;,
\label{kinetic}
\end{equation}
where $\bigtriangledown$ is the {\it del operator} and $\bigtriangleup$ is the {\it Laplace operator}.  Considering $H_{kin}\psi=E\psi$ and using the well known identities, 
\begin{eqnarray}
\bigtriangledown\left(fg\right)&=&f\bigtriangledown g+g\bigtriangledown f\nonumber\\
\bigtriangledown\cdot(f\vec{v})&=&f\bigtriangledown\cdot\vec{v}+\vec{v}\cdot(\bigtriangledown f)
\label{eq:id}\;,
\end{eqnarray}
the Schr\"odinger equation for a free quantum particle with variable mass can be written in a general way as
\begin{eqnarray}
-\frac{\hbar^2}{2m^*}\bigtriangleup\psi -\frac{\hbar^2}{2}\left[\bigtriangledown \left(\frac{1}{m^*}\right)\right]\bigtriangledown\psi -\frac{\hbar^2}{6}\left[\bigtriangleup \left(\frac{1}{m^*}\right)\right]\psi=E\psi \;.
\label{kinetic2}
\end{eqnarray}
Notice that if $m^*$ is constant, we have $-\frac{\hbar^2}{2m^*}\bigtriangleup\psi =E\psi$, and we recover the usual Schr\"odinger equation.  As mentioned above, we are interested in the study of quantum particles with variable mass constrained to move on curved surfaces. We follow the da Costa approach \cite{dacosta} in order to derive the general Schr\"odinger equation when $m^*$ depends on the position. It is important to point out that a refinement of the fundamental framework of the thin-layer quantization,  considering the surface thickness, has been carried out in \cite{Wang201668}. Nevertheless, we will not consider such  influence of the surface thickness here since the respective extra terms  must be treated using perturbation theory. This said, we consider a quantum particle constrained to move in a thin layer with constant width $d$, such that the constraint to the curved surface is achieved in the limit $d\rightarrow 0$. In  curvilinear coordinates $(q_1,q_2,q_3)$, where $q_1$ and $q_2$ are the longitudinal (tangent) coordinates while $q_3$ is the transverse (normal) one, the Schr\"odinger equation (\ref{kinetic2}) is written as
\begin{eqnarray}
&-&\frac{\hbar^2}{2m^*} \left[\frac{1}{\sqrt{G}}\partial_\mu\left(\sqrt{G}G^{\mu\nu}\partial_{\nu}\psi\right)\right]-\frac{\hbar^2}{2} \left[G^{\mu\nu}\left(\partial_\nu \frac{1}{m^*} \right) \left( \partial_\mu \psi \right) \right]\nonumber\\ 
&-&\frac{\hbar^2}{6} \left[\frac{1}{\sqrt{G}}\partial_\mu\left(\sqrt{G}G^{\mu\nu}\partial_{\nu}\frac{1}{m^*}\right)\right]\psi +V_\lambda (q_3)\psi = \imath \hbar \partial_t \psi\;.
\end{eqnarray}
Here, $G_{\mu\nu}$ are the coefficients of the {\it first fundamental form}, with the Greek letters $\mu,\nu=1,2,3$ and $G=\det(G_{\mu\nu})$. Following da Costa, we  added the potential $V_\lambda (q_3)$ which constrains the particle to the thin layer.  
In order to perform the separation of the longitudinal and transverse parts of the Schr\"odinger equation, we  decompose $G_{\mu,\nu}$ as
\begin{eqnarray}
G_{ij}&=&g_{ij}+\left[\alpha g+(\alpha g)^{T}\right]_{ij} q_3+\left(\alpha g\alpha^T\right)q_3^2\nonumber\\
G_{i3}&=&G_{3i}=0\nonumber\\
G_{33}&=&1\;,
\label{eq:gij}
\end{eqnarray}
where the Latin letters $i,j=1,2$ and $g_{ij}=\frac{\partial \vec{r}}{\partial q_i}\cdot\frac{\partial \vec{r}}{\partial q_j}$ is the induced metric on the surface.  The matrix $\alpha$ is the { Weingarten curvature operator} \cite{do2016differential}. The superscript $T$ denotes the  transposed matrix. It follows that the equations 
\begin{eqnarray}
&-&\frac{\hbar^2}{2m^*}\left(D \psi \right) - \frac{\hbar^2}{6}\left(D \frac{1}{m^*}\right)\psi - \frac{\hbar^2}{2}g^{ij}\left(\partial_j \frac{1}{m^*}\right)\left(\partial_i \psi \right) -\nonumber \\
&-& \frac{\hbar^2}{2}\left( \partial_{q_3}^2 \psi + \partial_{q_3} (\ln \sqrt{G})\partial_{q_3} \psi\right)\frac{1}{m^*} + V_\lambda (q_3)\psi = \nonumber \\
&&\imath \hbar \partial_t \psi,\;\;\;\;
\label{eqseparated}
\end{eqnarray}
where 
\begin{equation}
D = \frac{1}{\sqrt{g}}\partial_i \sqrt{g} g^{ij}\partial_j
\label{eq:lapbel}
\end{equation}
is the two-dimensional Laplace-Beltrami operator 
and  where we put $\partial_{q_3} \frac{1}{m^*}=\partial_{q_3}^2 \frac{1}{m^*} \equiv0$, since we are considering that the particle effective mass does not depend on the transversal coordinate $q_3$.

Writing the volume element as $dV=f(q_1, q_2, q_3)\sqrt{g}dq_1 dq_2 dq_3$, where $\sqrt{g}dq_1 dq_2$ is the area element of the surface and
\begin{equation}
f(q_1, q_2, q_3) = 1 + {\rm Tr}(\alpha_{ij})q_3 + \det(\alpha_{ij})q_3^2\;,
\end{equation}
suggests the wave function transformation
\begin{equation}
\chi(q_1, q_2, q_3) = [f(q_1, q_2, q_3)]^{1/2}\psi(q_1, q_2, q_3)\;, \label{transf}
\end{equation}
which allows for the definition of a surface probability density $|\chi_t (q_1, q_2)|^2 \int |\chi_n (q_3)|^2 dq_3$, in the event of a separation $\chi = \chi_t (q_1, q_2, t)  \chi_n (q_3, t)$.

Equation (\ref{eqseparated}), after substitution of \ref{transf} and taking $q_3 \rightarrow 0$, becomes
\begin{eqnarray}
&-&\frac{\hbar^2}{2m^*}  \left[\frac{1}{\sqrt{g}}\partial_i\left(\sqrt{g}g^{ij}\partial_{j}\chi\right)\right] -\frac{\hbar^2}{6}  \left[\frac{1}{\sqrt{g}}\partial_i\left(\sqrt{g}g^{ij}\partial_{j}\frac{1}{m^*}\right)\right]\chi\nonumber \\
&-&\frac{\hbar^2}{2} \left[g^{ij}\left(\partial_j \frac{1}{m^*} \right) \left( \partial_i \chi \right) \right]- \frac{\hbar^2}{2m^*}\left([\frac{1}{2}{\rm Tr}(\alpha_{ij})]^2 - \det(\alpha_{ij})\right)\chi\nonumber \\
&-&\frac{\hbar^2}{2m^*} \partial_{q_3}^2 \chi + V_\lambda (q_3)\chi = \imath \hbar \partial_t \chi\;.
\end{eqnarray}
Setting $\chi = \chi_t (q_1, q_2, t)  \chi_n (q_3, t)$, with the subscripts $t$ and $n$ meaning "tangent" and "normal", respectively, we can separate the above equation as follows:
\begin{eqnarray}
-\frac{\hbar^2}{2m^*} \partial_{q_3}^2 \chi_n + V_\lambda (q_3)\chi_n = \imath \hbar \partial_t \chi_n
\end{eqnarray}
and
\begin{eqnarray}
&-&\frac{\hbar^2}{2m^*} \left[\frac{1}{\sqrt{g}}\partial_i\left(\sqrt{g}g^{ij}\partial_{j}\chi_t \right)\right]-\frac{\hbar^2}{6}\left[\frac{1}{\sqrt{g}}\partial_i\left(\sqrt{g}g^{ij}\partial_{j}\frac{1}{m^*}\right)\right]\chi_t \nonumber\\
&-&\frac{\hbar^2}{2} \left[g^{ij}\left(\partial_j \frac{1}{m^*} \right) \left( \partial_i \chi_t \right) \right]-\frac{\hbar^2}{2m^*}\left([\frac{1}{2}{\rm Tr}(\alpha_{ij})]^2 - \det(\alpha_{ij})\right)\chi_t\nonumber\\&=& \imath \hbar \partial_t \chi_t\;.\label{long}
\end{eqnarray}
The last term on the left-hand side of Eq.  (\ref{long}) is the geometric potential, that we will call {\it  da Costa potential}, $V_{\rm daCosta}=- \frac{\hbar^2}{2m^*}\left([\frac{1}{2}{\rm Tr}(\alpha_{ij})]^2 - \det(\alpha_{ij})\right)$, which can be written as  
\begin{equation}
V_{\rm daCosta}=- \frac{\hbar^2}{2m^*}\left[\rm M^2-K\right]\;, \label{da}
\end{equation}
where ${\rm M}$ and ${\rm K}$ are, respectively, the mean and Gaussian curvature of the given surface. Note that the only change in the potential as compared to the constant mass case is  that now $m^*\equiv m^*(q_1,q_2)$. On the other hand,  the Schr\"odinger equation now has two extra pieces which contain  derivatives of the effective mass with respect to the particle's position. Da Costa's original Schr\"odinger equation is recovered when the mass is made constant. 

It is important to mention that we are considering here the simplest case for the effective mass, which appears in systems with an isotropic dispersion relation. For systems with non-symmetric energy bands, the term $1/m^*$ must be replaced by the effective mass tensor, which in the diagonalized form is given by
\begin{eqnarray}
\left[\frac{1}{m^*}\right]^{\mu \nu} = 
\left(
\begin{array}{ccc}
m^{-1}_{11} & 0 & 0 \\
0 & m^{-1}_{22} & 0 \\
0 & 0 & m^{-1}_{33}
\end{array}
\right),
\end{eqnarray}
where $m_{11}$, $m_{22}$ and $m_{33}$ are the principal masses. A study of the influence of an anisotropic effective mass in the transport properties of 2D quantum systems is presented in \cite{souza2018curved}.

Eq. (\ref{long}) is the Schr\"odinger equation for a quantum particle confined to a generic surface with a position-dependent mass. In the next section, we will consider the special case of a surface of revolution, obtaining the transmittance and reflectance for these surfaces in terms of the boundary conditions.

\section{Surfaces of Revolution}

Let us identify   $q_2$ with the Cartesian coordinate $y$. The rotation of a planar   curve $\rho(q_2)$ about the axis $y$ yields a surface of revolution  parametrized  by
\begin{eqnarray}
x&=&\rho(q_2)\cos(\varphi),\nonumber \\ 
y&=&q_2, \\
z&=&\rho(q_2)\sin(\varphi). \nonumber
\end{eqnarray}
 $\rho(q_2)$ is therefore the cylindrical radius of the surface at each point.

The first and second fundamental form for surfaces of revolution are given by
\begin{equation}
\textbf{g}_{ij}=\left(
\begin{array}{cc}
\rho(q_2)^2 & 0 \\
0 & 1+\rho^\prime(q_2)^2 \\    
\end{array} \right)
\label{fff}
\end{equation}
and
\begin{equation}
\textbf{h}_{ij}= \frac{1}{\sqrt{1+\rho^\prime(q_2)^2}}\left(
\begin{array}{cc}
\rho(q_2) & 0 \\
0 & -\rho^{\prime\prime}(q_2) \\   
\end{array} \right),
\end{equation}
respectively. Since the mean and Gaussian curvatures are, respectively,
\begin{equation}
M=\frac{1}{2g}(g_{11}h_{22}+g_{22}h_{11}-2g_{12}h_{12})
\end{equation}
and
\begin{equation}
K=\frac{1}{g}(h_{11}h_{22}-h_{12}h_{21}),
\end{equation}
we have that
\begin{equation}
M=\frac{1+\rho^\prime(q_2)^2-\rho(q_2)\rho^{\prime\prime}(q_2)}{2\rho(q_2)[1+\rho^\prime(q_2)^2]^{3/2}},
\end{equation}
and
\begin{equation}
K=-\frac{\rho^{\prime\prime}(q_2)}{\rho(q_2)[1+\rho^\prime(q_2)^2]^2}.
\end{equation}
With the above results for $M$ and $K$ substituted in Eq. \ref{da}, we obtain the da Costa potential for a surface of revolution as
\begin{equation}
V_{daCosta}=-\frac{\hbar^2}{8m^*}\frac{[1+\rho^\prime(q_2)^2+\rho(q_2)\rho^{\prime\prime}(q_2)]^2}{\rho(q_2)^2[1+\rho^\prime(q_2)^2]^3}.
\label{psr}
\end{equation}

For a surface of revolution, the effective mass does not depend on the angular coordinate $\varphi$. Therefore, the da Costa potential depends only on the coordinate $q_2$. It allows us to use separation of variables in the Schr\"odinger equation. Then, with the first fundamental form (\ref{fff}) and the potential (\ref{psr}), it is possible to separate Eq. (\ref{long}) into      

\begin{equation}
\Phi^{\prime\prime} + A\Phi = 0
\label{phi}
\end{equation}
and
\begin{eqnarray}
\Upsilon^{\prime\prime} + [F(q_2) + m(m^{-1})^\prime]\Upsilon^\prime + [m/3((m^{-1})^{\prime\prime}\nonumber \\
 + F(q_2)(m^{-1})^\prime) + G(q_2)(E_I - V)2m/\hbar^2]\Upsilon = 0,
 \label{q}
\end{eqnarray}
where $\Phi(\varphi)$ and $\Upsilon(q_2)$ are the angular and axial eigenfunctions, respectively, $A$ is the separation constant,
\begin{equation}
F(q_2)= \rho^{-1}\rho^\prime[1-\rho\rho^{\prime\prime}(1+(\rho^\prime)^2)^{-1}]
\end{equation}
and
\begin{equation}
G(q_2)= 1+(\rho^\prime)^2.
\end{equation}
If the mass is constant we retrieve the equations in \cite{Marchi}.

Recalling that Eq. (\ref{q}) is of the form
\begin{equation}
\Psi(y)^{\prime\prime} + V_1 \Psi(y)^{\prime}  + V_2 \Psi(y) = 0,
\label{Psieq}
\end{equation}
by making $\Psi(y)=\phi(y)\lambda(y)$, we get
\begin{equation}
\phi^{\prime\prime} + \left(2\frac{\lambda^{\prime}}{\lambda} + V_1 \right) \phi^{\prime}  + \left(\frac{\lambda^{\prime\prime}}{\lambda}+ V_1 \frac{\lambda^{\prime}}{\lambda} +V_2 \right) \phi = 0.
\end{equation}

Now, with 
\begin{equation}
2\frac{\lambda^{\prime}}{\lambda} + V_1 =0
\label{lambdaeq}
\end{equation} and 
\begin{equation}
\frac{\lambda^{\prime\prime}}{\lambda}+ V_1 \frac{\lambda^{\prime}}{\lambda} +V_2 = W ,
\end{equation}
we have
\begin{equation}
\phi^{\prime\prime}  + W \phi = 0,
\end{equation}
with $\lambda(y)$ given by the solution of Eq. (\ref{lambdaeq}):
\begin{equation}
\lambda(y) = e^{-\frac{1}{2}P(y)}, \label{lambdares}
\end{equation}
where $P(y)$ is the primitive for the function $V_1(y)$, such that $P^{\prime}(y)=V_1(y)$.  With this, $W=-\frac{1}{4} V_1^2 - \frac{1}{2} V_1^{\prime} + V_2$. The solution to Eq. (\ref{Psieq}) may then be found by solving
\begin{equation}
\phi^{\prime\prime}  + \left( -\frac{1}{4} V_1^2 - \frac{1}{2} V_1^{\prime} + V_2 \right) \phi = 0. \label{phieq}
\end{equation}

We are interested in understanding the effective mass effect in the transmittance and reflectance of incident electrons in a surface of revolution with length $L$ and a specific $\rho(y)$. We consider the injection of electrons of energy $E_k$ coming from the negative part of the $y$-axis. Thus, we have
\begin{align}
\Psi (y) & =  a_0 e^{ik_0 y} + b_0 e^{-i k_0 y} \; \; \text{for} \, y \leq 0,\ \nonumber \\ 
            & = a_L e^{-ik_L (y-L)} + b_L e^{i k_L( y-L)} \; \;  \text{for}\, y \geq L, \label{psiinout}
\end{align}
where 
\begin{equation}
k_0 = \sqrt{\frac{2m^{*}(0)}{\hbar^2}(E_k-V(0))} \label{k0}
\end{equation}
and
\begin{equation}
k_L= \sqrt{\frac{2m^{*}(L)}{\hbar^2}(E_k-V(L))} \label{kL}
\end{equation}
are the incident and transmitted electron wavectors, respectively.  Notice that  $V_1(y)=0$ for $y$ not in the range $0<y<L$, which makes Eqs. (\ref{Psieq}) and (\ref{phieq}) identical. So, Eq. (\ref{psiinout}) is also valid for $\phi(y)$ and then $\Psi(0)=\phi(0)$ as well as $\Psi(L)=\phi(L)$. As we will see bellow, this implies that the reflectance and transmittance do not depend on $\lambda(y)$ and can be obtained from $\phi(y)$ directly. We choose the normalization of the incident wavefunction such that $a_0=1$. Also, considering only outgoing waves in the $y\geq 0$ region, we have  $a_L =0$. These are the Robin boundary conditions for the problem \cite{Marchi,santos}.

For the above normalization, the transmittance and reflectance may be obtained from the probability current density
\begin{equation}
j=\frac{\hbar}{2mi}\left( \Psi^* \Psi^{'}- \Psi \Psi^{* }\right),
\end{equation}
such that the incident current is
\begin{equation}
j_{inc}= \frac{\hbar k_0}{m^{*}(0)},
\end{equation}
the reflected current is
\begin{equation}
j_{ref} = \frac{\hbar k_0}{m^{*}(0)} |b_{0}|^2
\end{equation}
and the transmitted current
\begin{equation}
j_{trans}= \frac{\hbar k_L}{m^{*}(L)} |b_L|^2 ,
\end{equation}
all in absolute values.

This way, the transmittance is
\begin{equation}
T=\frac{j_{trans}}{j_{inc}}=\frac{m^{*}(0)}{m^{*}(L)}\frac{k_L}{k_0}| b_L |^2 \label{trans}
\end{equation}
and the reflectance,
\begin{equation}
R=\frac{j_{ref}}{j_{inc}}=| b_0 |^2 .
\end{equation}

In terms of the wavefunction, the boundary conditions $a_0=1$ and $a_L =0$  are
\begin{equation}
a_0 = \frac{1}{2}\left[ \phi(0) - i \phi^{\, '}(0)/k_0 \right]=1 \label{a0}
\end{equation}
and
\begin{equation}
a_L = \frac{1}{2}\left[ \phi(L) + i \phi^{\, '}(L)/k_L \right]=0 , \label{aL}
\end{equation}
since $\Psi(y)=\phi(y)$ in $y=0$ and $y=L$. From (\ref{psiinout}) we also have that
\begin{equation}
b_0=\phi(0)-1
\end{equation}
and
\begin{eqnarray}
b_L   =  \phi(L)  . \label{bLphi}
\end{eqnarray}
It folows that 
\begin{equation}
T=\frac{m^{*}(0)}{m^{*}(L)}\frac{k_L}{k_0} \left| \phi(L) \right|^2  \label{transnew}
\end{equation}
and
\begin{equation}
R=|\phi(0)-1 |^2 \label{refnew} .
\end{equation}
Then, the problem reduces to finding $\phi(0)$ and $\phi(L)$, which is done by solving the coupled differential and algebraic equations (\ref{phieq}), (\ref{a0}) and (\ref{aL}) in the range $0\leq y\leq L$. This is the essence of the open boundary condition method for solving ordinary differential equations with Robin boundary conditions, \textit{i.e.}, the specification of a linear combination of the values of a function and the values of its derivative on the boundary of the domain.

As a first application for the above results, in the next section we will investigate the transport properties of deformed nanotubes with a position dependent mass.

\section{Deformed Nanotubes}

Let us now obtain the transport properties of deformed nanotubes. We will consider corrugated nanotubes, with the corrugation generated by the curve
\begin{equation}
\rho(y)= r \left[ 1 + \frac{\epsilon}{2}\left( 1 - \textrm{cos}\left( 2 \frac{n\pi y}{L} \right) \right)\right],
\label{rho}
\end{equation}
where $r$ is the initial radius of the nanotube, $\epsilon$ is a parameter that gives the strength of the increase (for positive values) or decrease (for negative values) of the radius of the nanotube, $L$ is the length of the corrugated region and $n$ gives the number of corrugations. This problem was recently addressed both for corrugations in the plane \cite{ono2009tuning} and in nanotubes \cite{santos}, but the variation of the effective mass with the position was neglected. 

We will consider the empirical dependence of the effective mass with the radius of the nanotube that was obtained in Ref. \cite{5175416} for carbon nanotubes
\begin{equation}
m^{*}= \frac{8m_{e}\hbar^{2}}{9\gamma^{2}_{0}a_{cc}}d^{-0.7835}e^{7.4\times 10^{-3}\theta},
\end{equation}
where $m_e$ is the mass of free electrons, $d$ is the diameter of the nanotube, $\gamma_{0}=2.44\textrm{eV}$ is the overlap energy and $a_{cc}$ is the first neighbor distance between the carbon atoms. $\theta$ is the chiral angle and gives the type of the carbon nanotube. We will consider $\theta=0$, which means a metallic zigzag carbon nanotube.
It is important to mention that in order to input the energy in meV and distances in nm we use a mixed units system where the electron mass is $m_e = 5.68 \times 10^{-27}$  meV.s$^2$/nm$^2$ and Planck's constant is $\hbar= 6.58 \times 10^{-13}$ meV.s.

\begin{figure*}[h!]
 
  \centering
   \includegraphics[angle=0,width=4 cm,height=4.5 cm ]{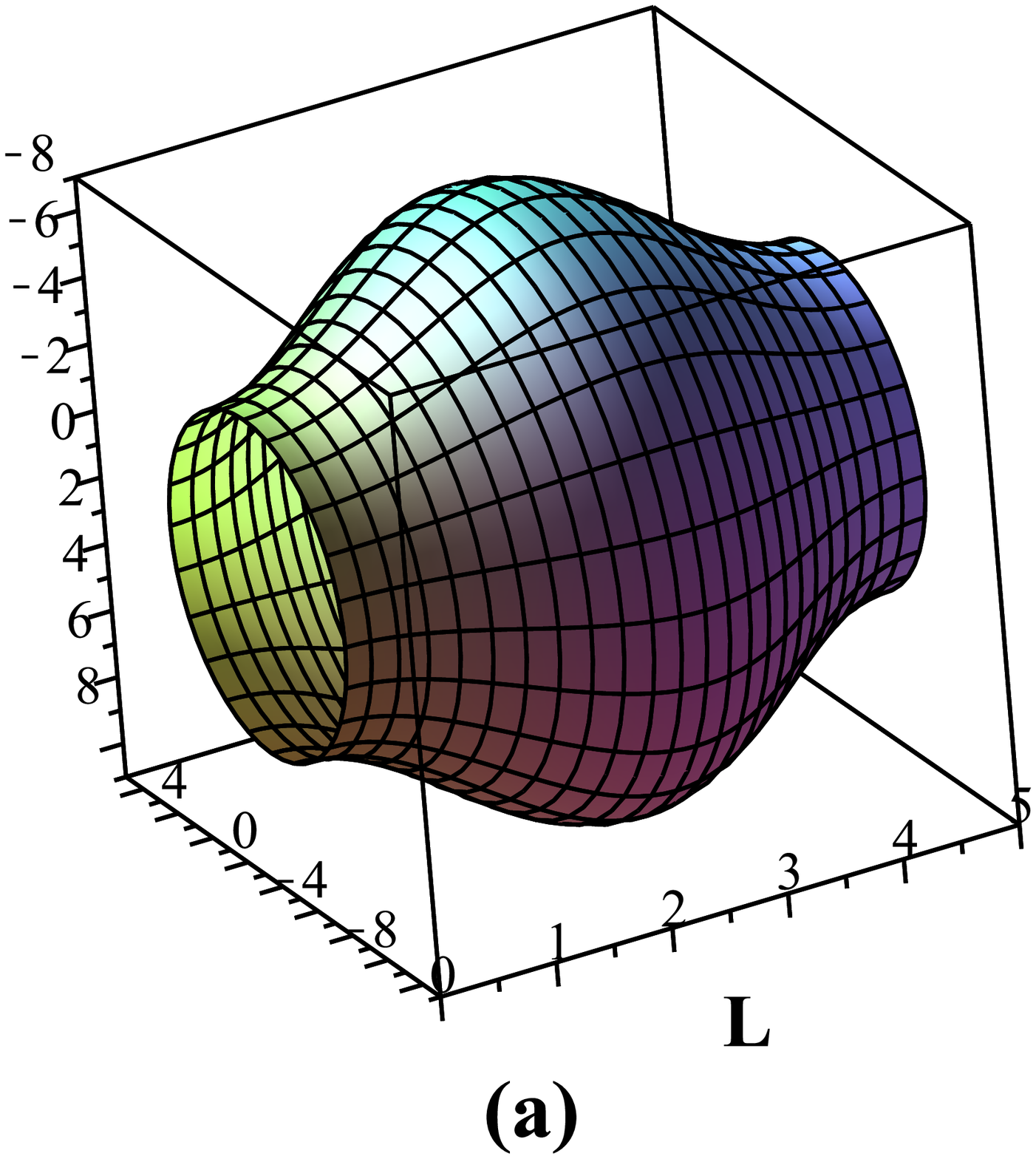}
\includegraphics[angle=0,width=6 cm,height=4.5 cm ]{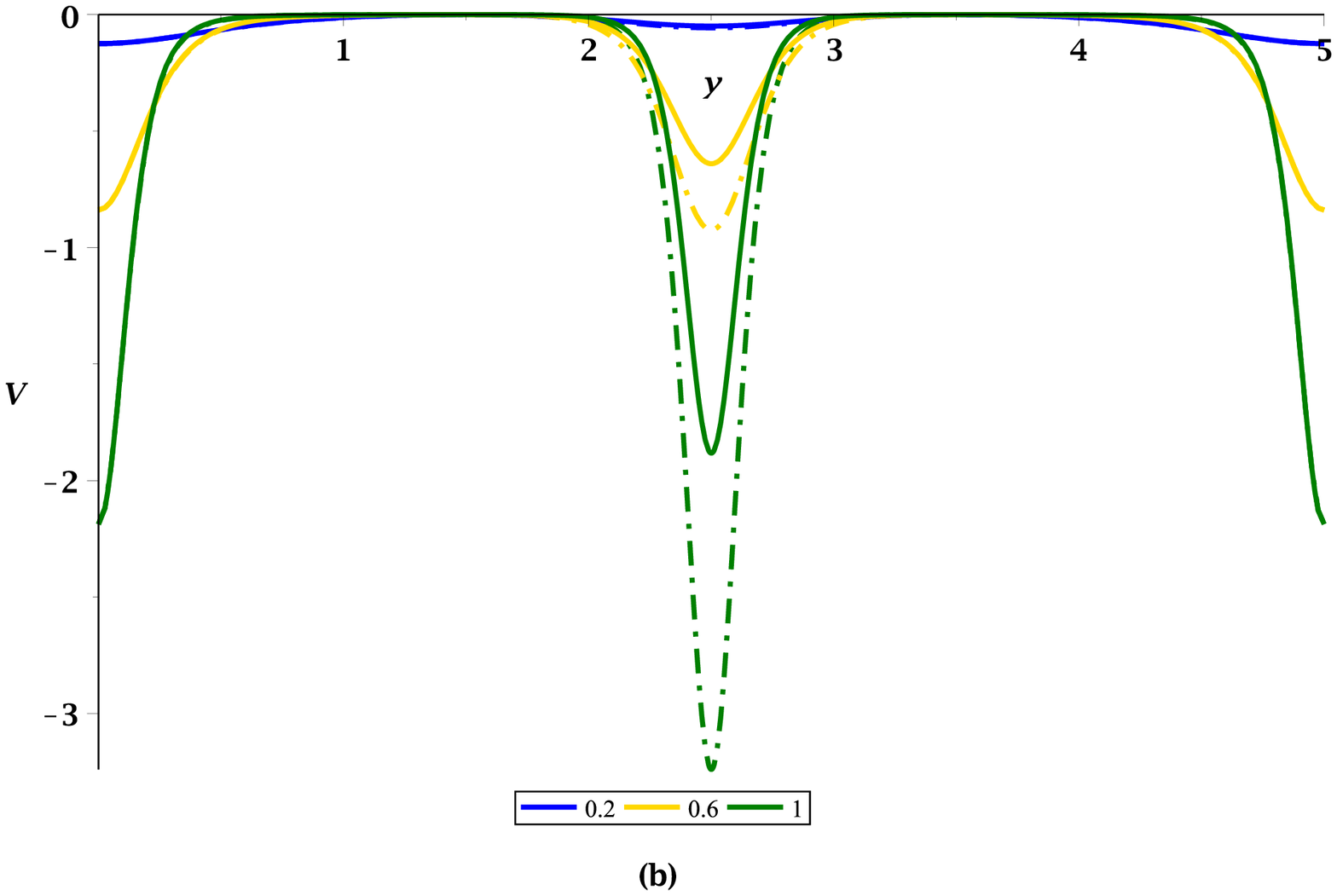}
\includegraphics[angle=0,width=6 cm,height=4.5 cm ]{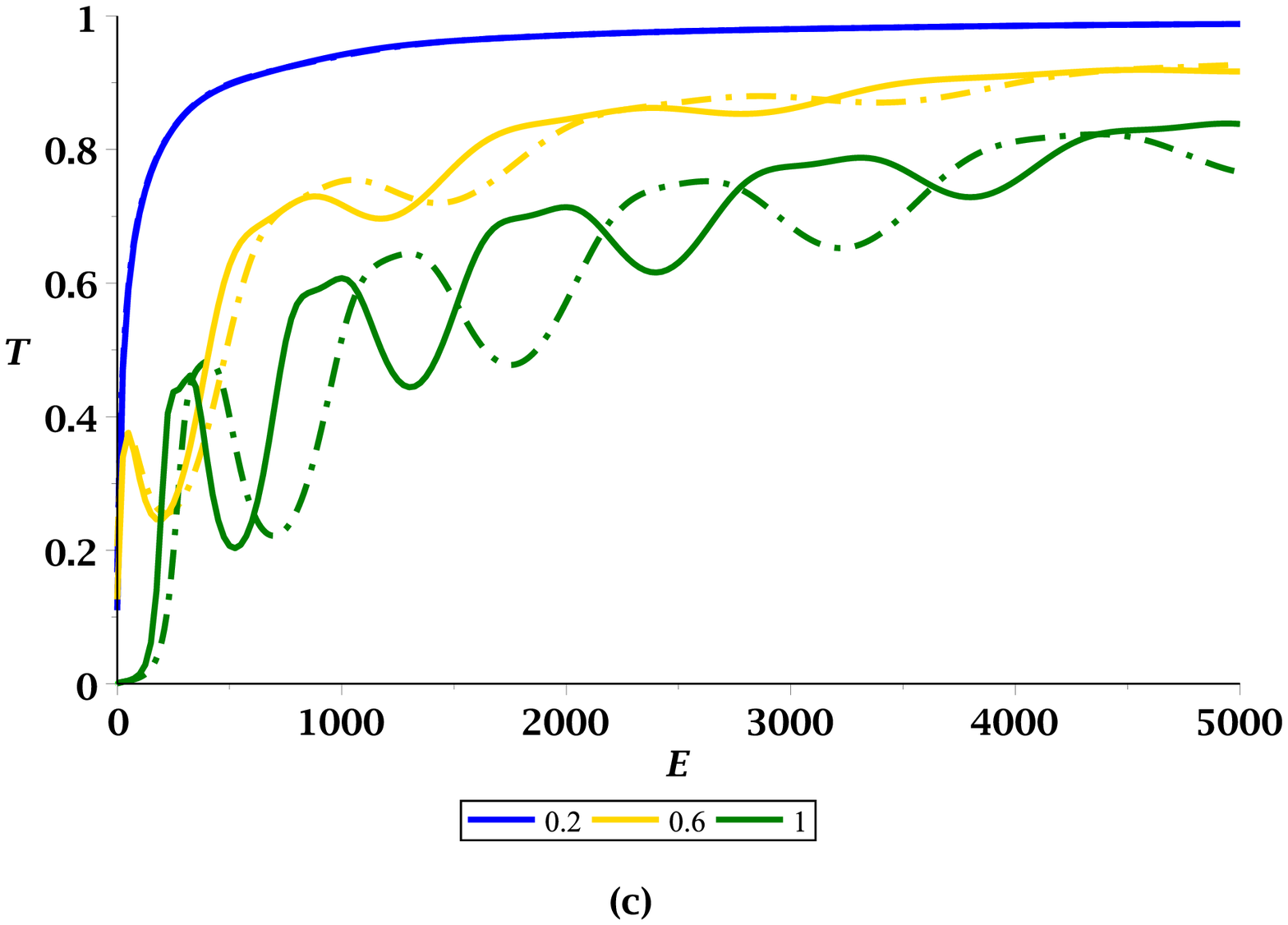}
     \caption{The geometric potential (b) and the transmittance (c) with different values of $\epsilon$ for the case of one bump in the nanotube, as shown in (a). The continuum lines represent the constant mass case, while the dash-dotted lines the variable mass case.}
     \label{1bump}
\end{figure*}

\begin{figure*}[h!]

  \centering
\includegraphics[angle=0,width=4 cm,height=4.5 cm ]{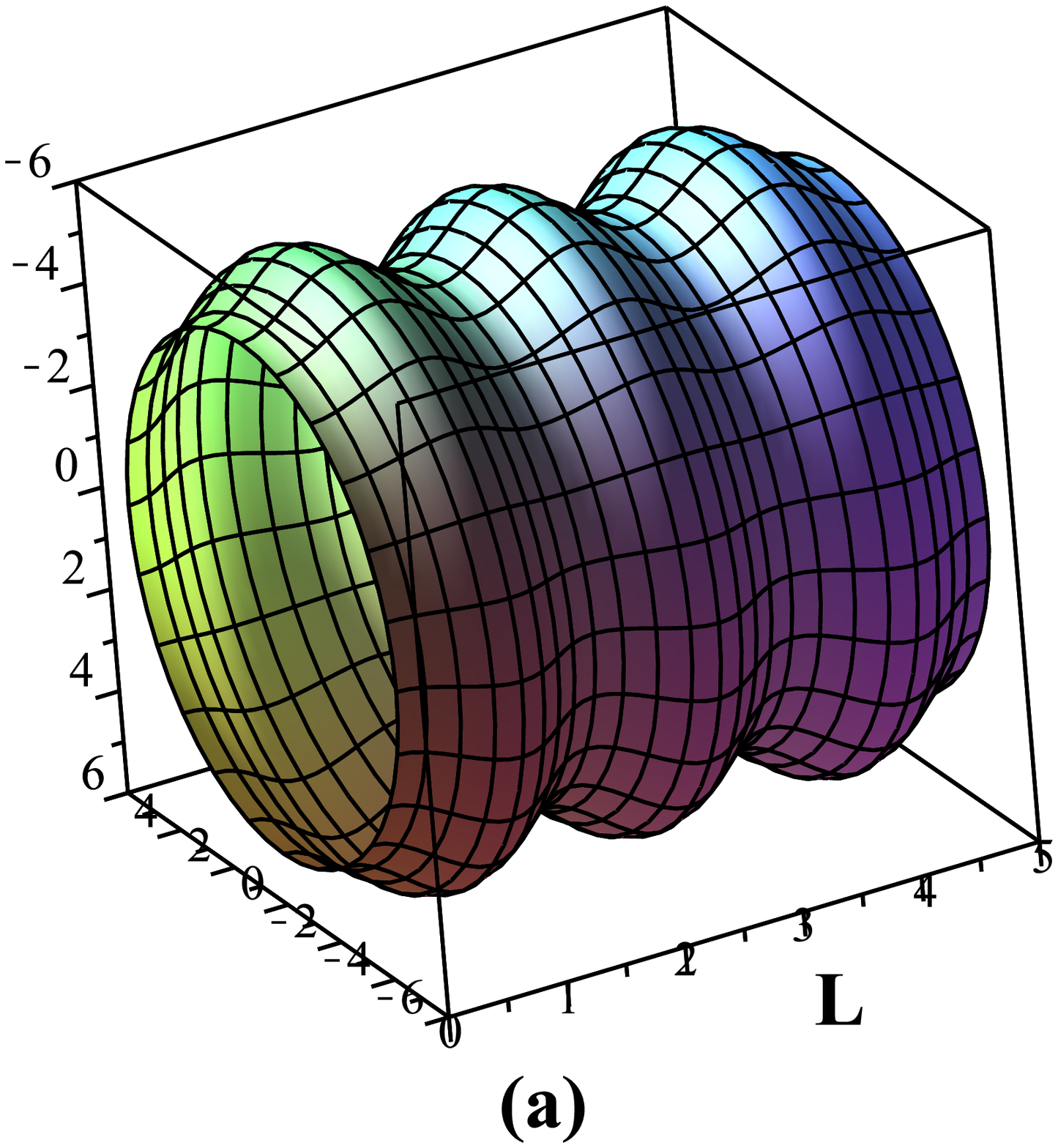}
\includegraphics[angle=0,width=6 cm,height=4.5 cm ]{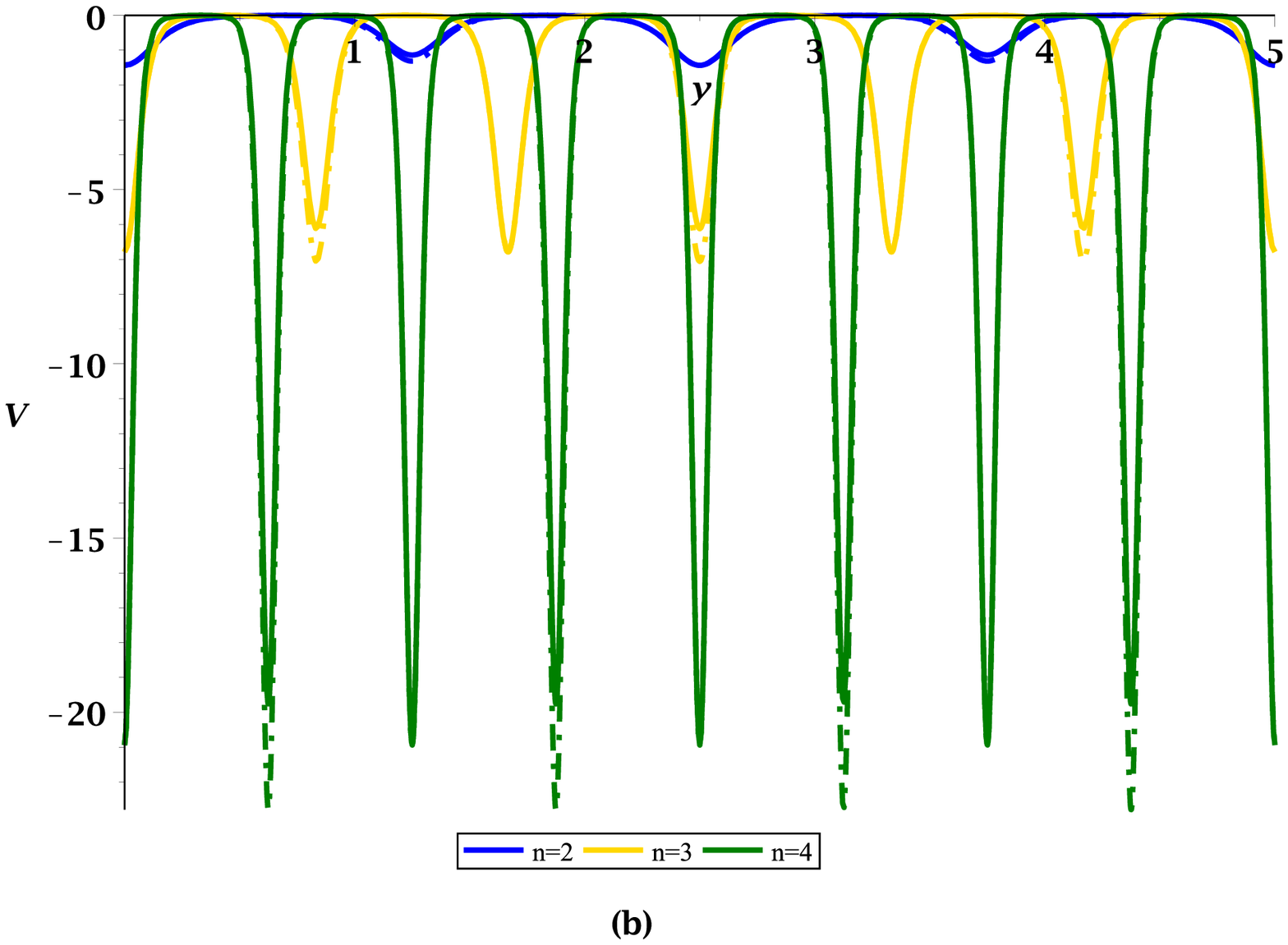}
\includegraphics[angle=0,width=6 cm,height=4.5 cm ]{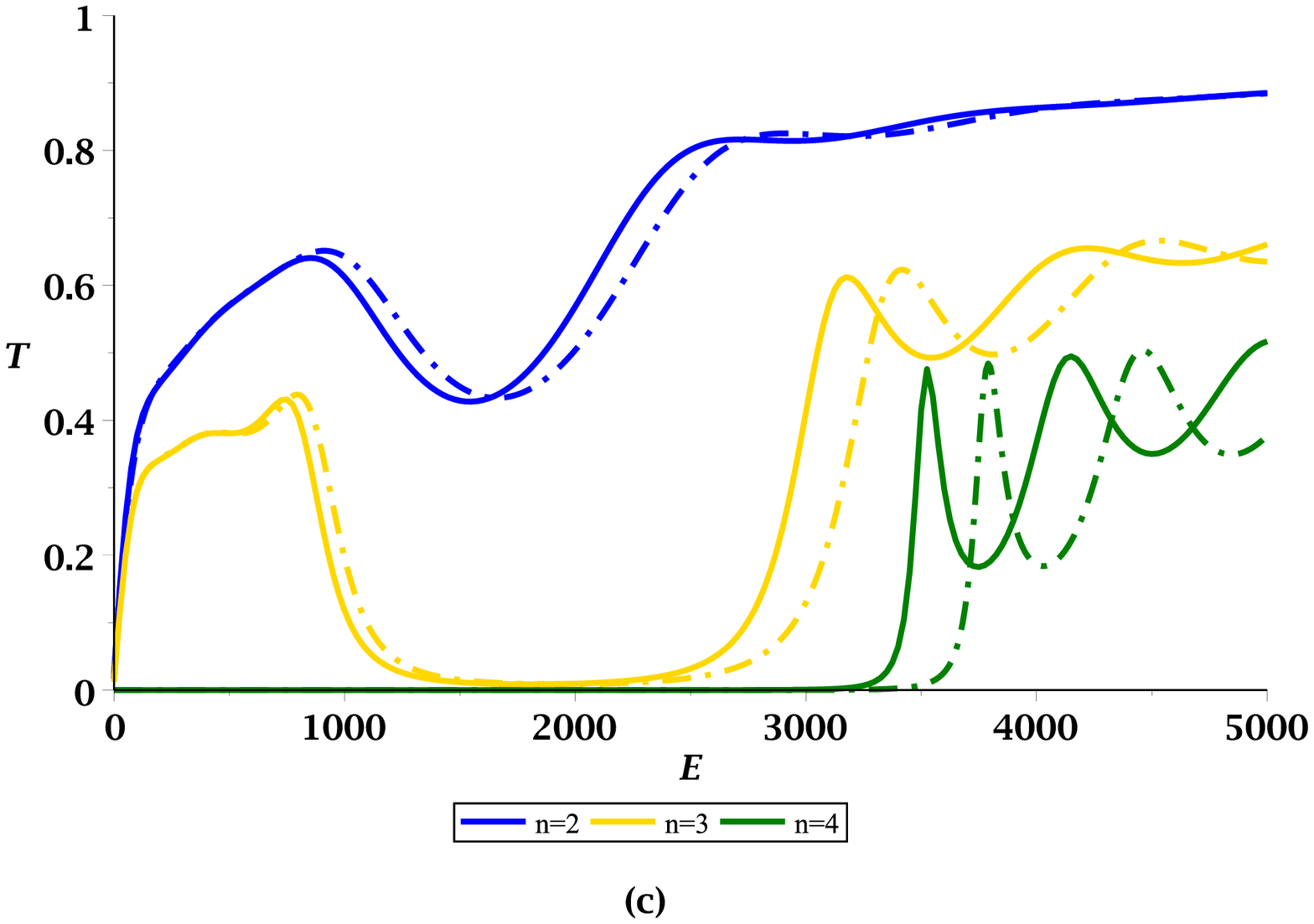}
     \caption{The geometric potential (b) and the transmittance (c) with $\epsilon = 0.2$ for different values of $n$, i. e., $n$ bumps in the nanotube, as shown in (a). The continuum lines represent the constant mass case, while the dash-dotted lines the variable mass case.}
     \label{nbump}
\end{figure*}

\begin{figure*}[h!]
 
  \centering
 \includegraphics[angle=0,width=4 cm,height=4.5 cm ]{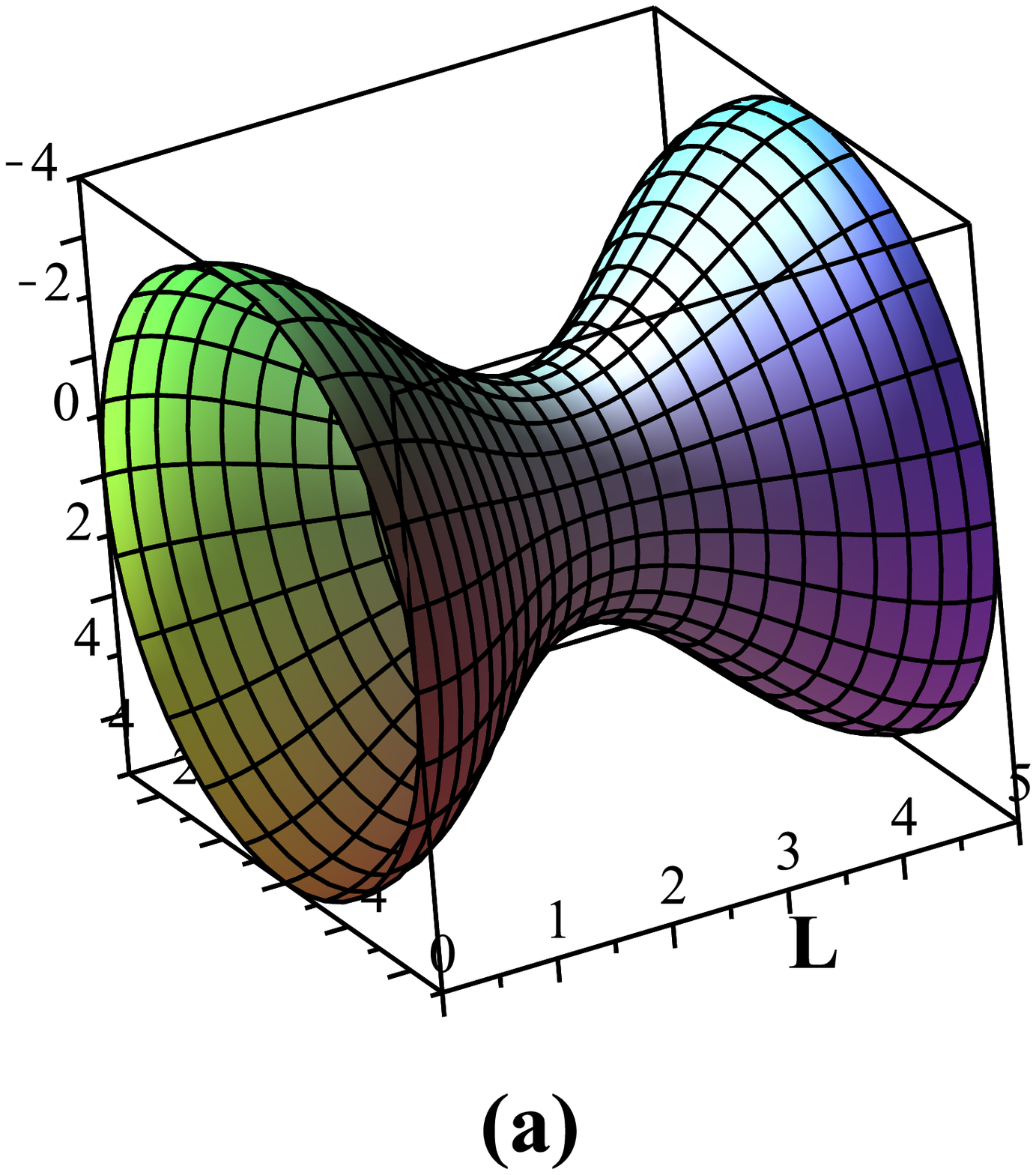}
\includegraphics[angle=0,width=6 cm,height=4.5 cm ]{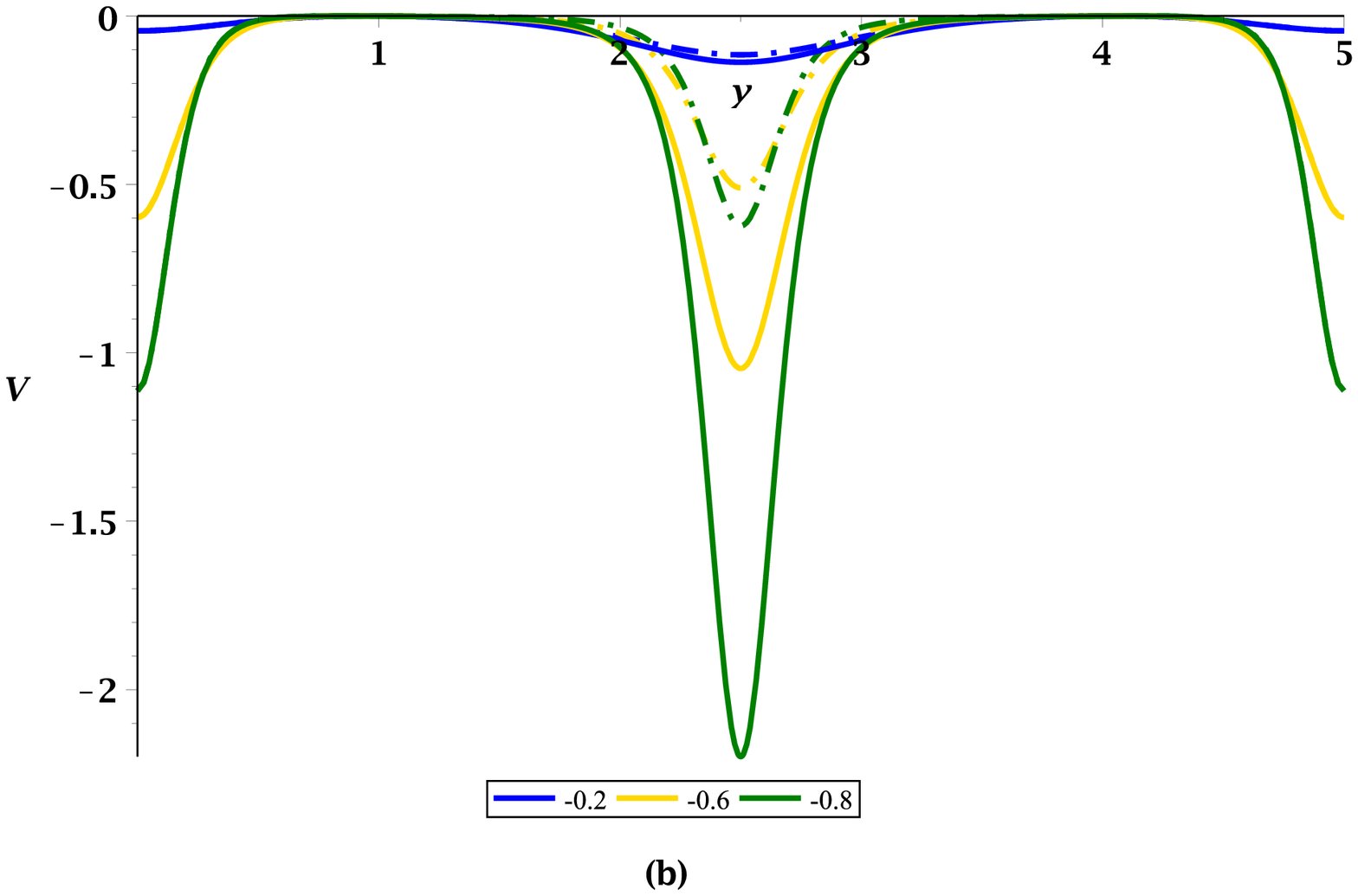}
\includegraphics[angle=0,width=6 cm,height=4.5 cm ]{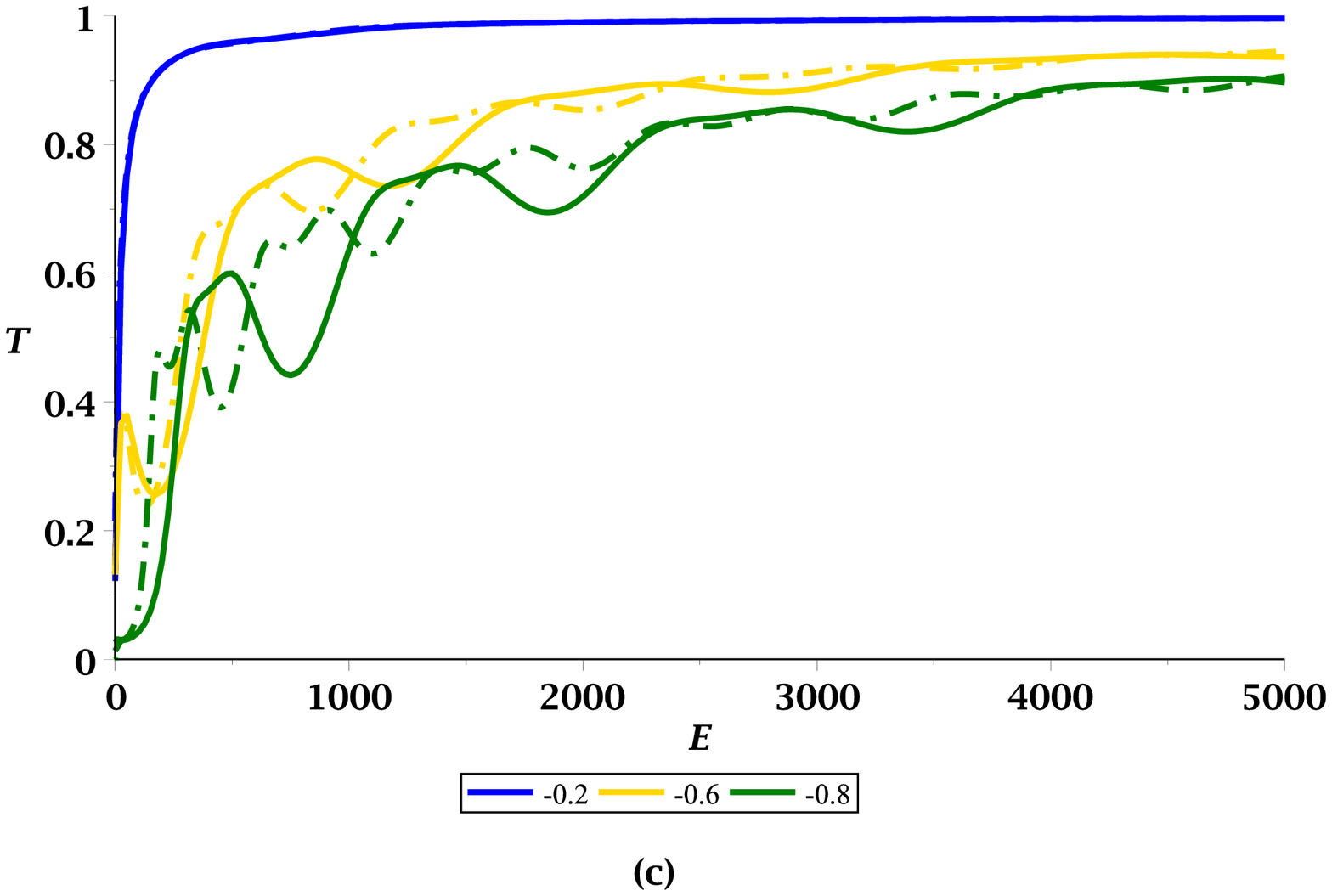}
     \caption{The geometric potential (b) and the transmittance (c) with different values of $\epsilon$ for the case of one depression in the nanotube, as shown in (a). The continuum lines represent the constant mass case, while the dash-dotted lines the variable mass case.}
     \label{1dep}
\end{figure*}

\begin{figure*}[h!]
 
  \centering
 \includegraphics[angle=0,width=5 cm,height=4.5 cm ]{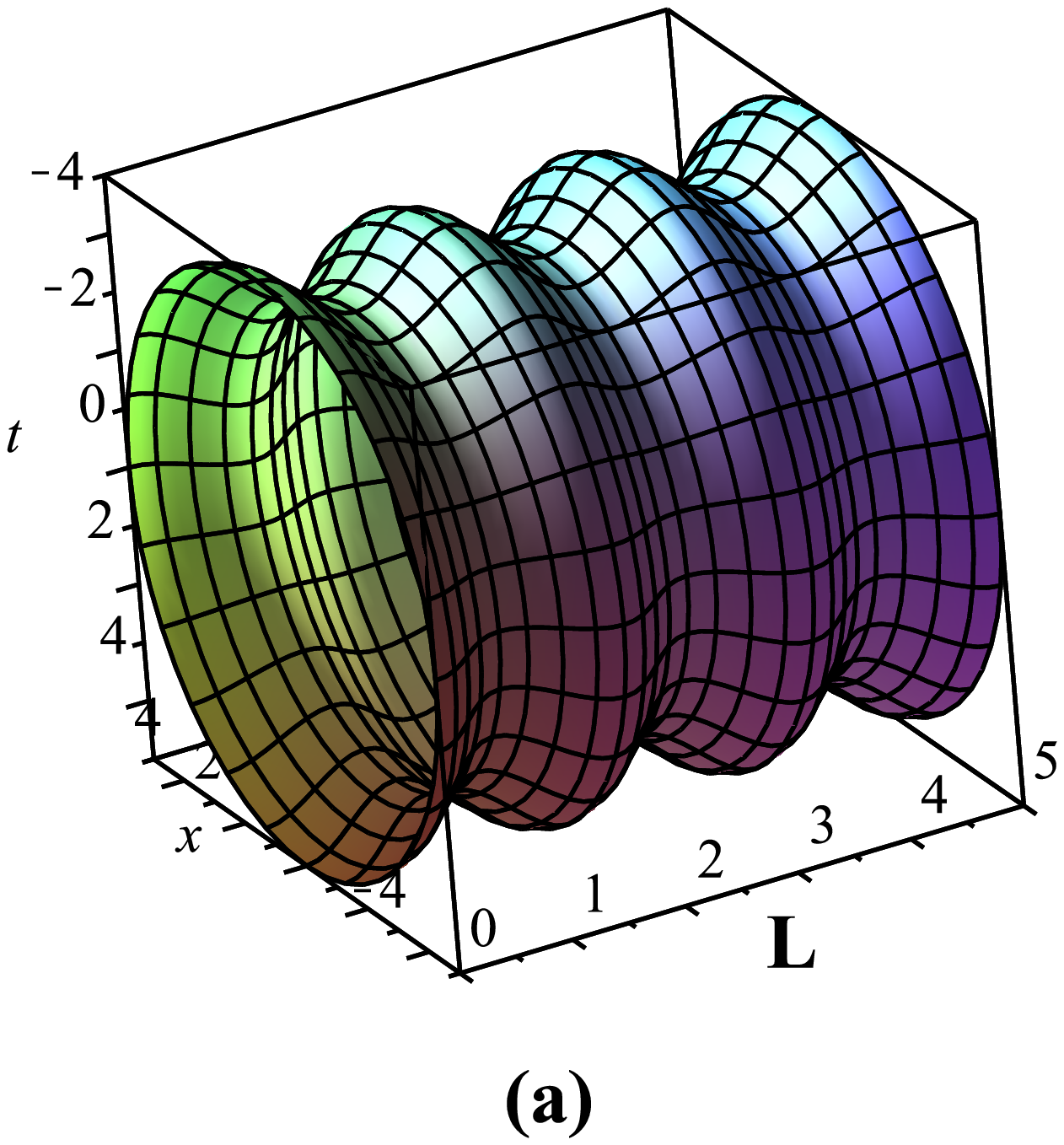}
\includegraphics[angle=0,width=6 cm,height=4.5 cm ]{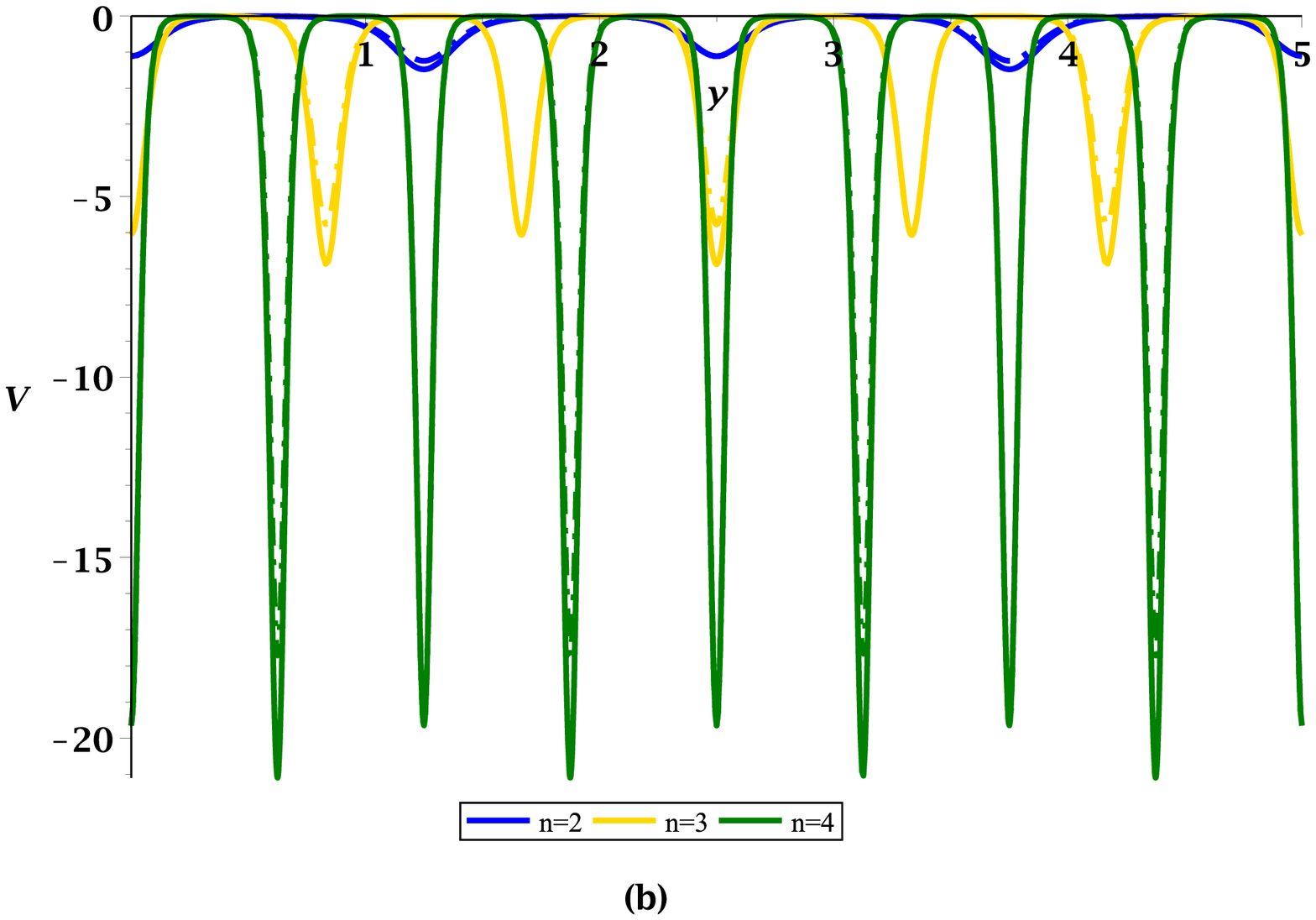}
\includegraphics[angle=0,width=6 cm,height=4.5 cm ]{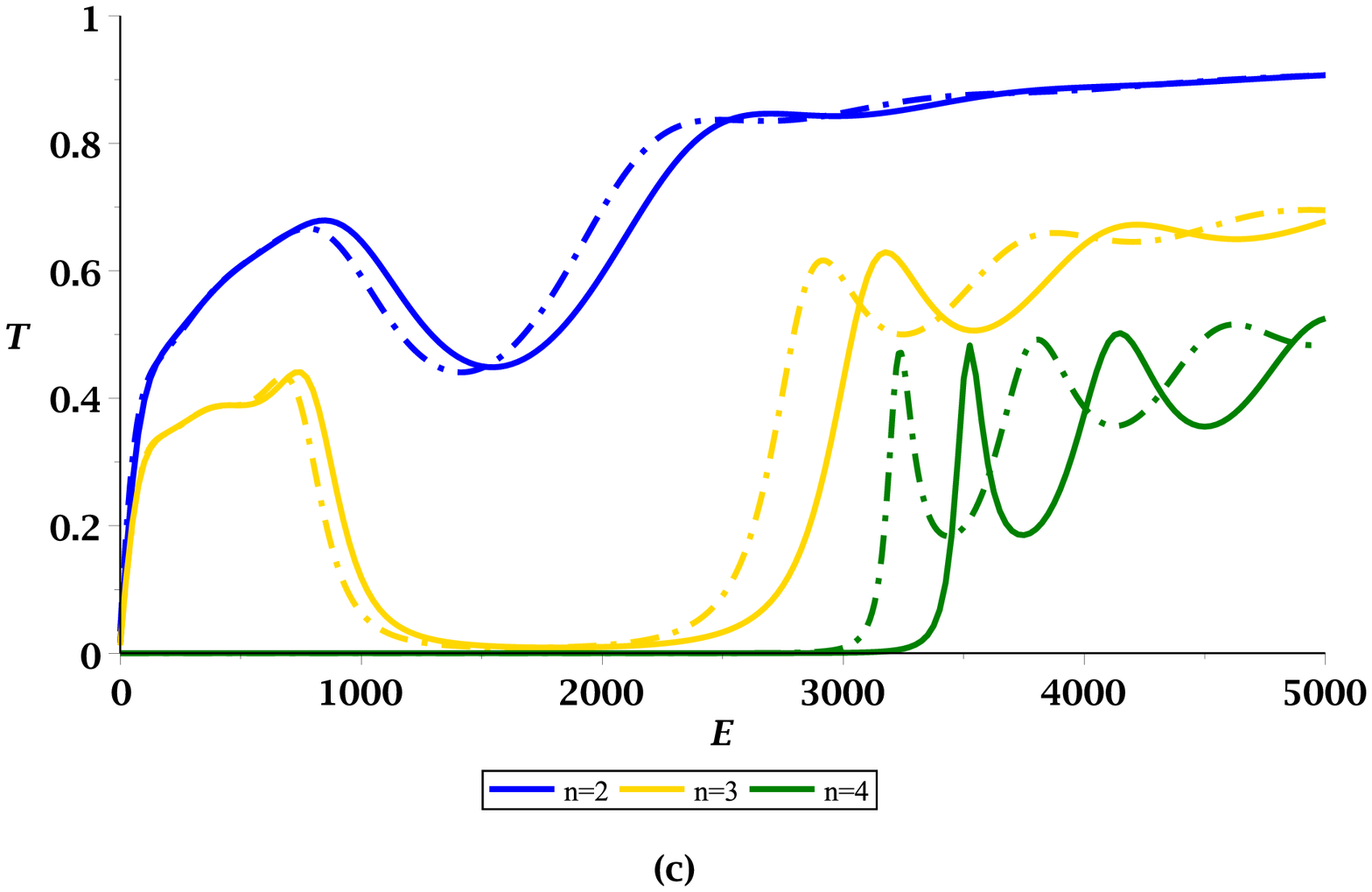}
     \caption{The geometric potential (b) and the transmittance (c) with $\epsilon = -0.2$ for different values of $n$, i. e., $n$ depressions in the nanotube, as shown in (a). The continuum lines represent the constant mass case, while the dash-dotted lines the variable mass case.}
     \label{ndep}
\end{figure*}

We implemented a MAPLE code to solve recursively the mixed set of  differential and algebraic equations (\ref{phieq}), (\ref{a0}) and (\ref{aL})  and find, for each injection energy, $\phi(0)$  and $\phi(L)$ and, consequently, the transmittance and the reflectance as specified by Eqs. (\ref{transnew}) and (\ref{refnew}), respectively. The results are in Figs. \ref{1bump} to \ref{ndep}, where we consider the constant (continuum lines) and variable (dashed dotted lines) mass cases in order to analyze the effects of the position depend mass in the transport properties of the deformed nanotubes. For all cases considered here, $r$ in Eq. (\ref{rho}) is equal to $5$ nm. In the constant mass case, we consider that the mass does not change in the deformation, which means to consider the effective mass of a nanotube with $\epsilon=0$ in Eq. (\ref{rho}). The effective mass for this case can be seen in Table 1, where we define the parameter $\zeta = m^*/m_e$, which gives the strength of the modification of the effective mass compared to the mass of free electrons.

In Fig. \ref{1bump} we consider that the deformation of the nanotube is given by one bump, as shown in Fig. \ref{1bump} $(a)$. The values of $\zeta$ in the middle of each bump considered can be found in Table 1. As was already discussed here, the PDM does not modify the form of the geometric potential, only  its intensity. In fact, the potential is inversely proportional to the effective mass. Since an increase of the radius of the nanotube induces a decrease in the effective mass, as can be seen in Table 1, the variable mass case will have a deeper geometrical potential in comparison with the constant mass case. This is shown in Fig. \ref{1bump} $(b)$. It is possible to note that, as the value of $\epsilon$ increases, which means a  higher bump, the inclusion of the PDM becomes more important.  This can be seen clearly in Fig. \ref{1bump} $(c)$. One can see that the PDM alters the values of energy where the resonant peaks in the transmittance occur, revealing the importance of considering the PDM in the transport properties of the system. It is important to remember that the PDM introduces two extra terms in the Hamiltonian, besides its modification of the geometrical potential. 

In Fig. \ref{nbump} we consider a nanotube with more than one bump.  For a single {small}   (e.g., $\epsilon=0.2$) bump there is not much difference between the constant and variable mass cases, {as can be seen in the blue lines in Fig. \ref{1bump}}. However, as the number of bumps increases, the influence of the PDM becomes more and more important. In fact, increasing the number of bumps, the geometric potential tends to a Dirac comb, as can be seen in Fig. \ref{nbump} $(b)$. As a consequence, energy gaps appear in the energy spectrum. These energy gaps become better defined as the number of bumps increases. The inclusion of the PDM corrects the width of the energy gap and its location in the energy spectrum, as  shown in Fig. \ref{nbump} $(c)$. In fact, the energy gap for the variable mass case is larger as compared with the constant mass case. It reveals that, in order to use curvature to design nanotube-based electronic devices, the PDM must be considered. 

\begin{table}[]
\centering
\caption{The values of the parameter $\zeta$ in the center of the deformation for different values of $\epsilon$.}
\label{my-label}
\begin{tabular}{|c|c|}
\hline
\textbf{$\epsilon$} & \textbf{$\zeta (m^*/m_e)$} \\ \hline
1                   & 0.04353858056    \\ \hline
0.6                 & 0.05185650718    \\ \hline
0.2                 & 0.06496699234    \\ \hline
0                   & 0.07494303341	   \\ \hline
-0.2                & 0.08926069477    \\ \hline
-0.6                & 0.1536445870     \\ \hline
-0.8                & 0.2644686912     \\ \hline
\end{tabular}
\end{table}

The case of a pinched nanotube is considered in Fig. \ref{1dep}. In contrast with the case of a bump, as the radius of the nanotube decreases, the effective mass increases, which induces a shallower geometrical potential as compared with the constant mass case.  As in the case of bumps, the PDM becomes more relevant as the value of $\epsilon$ decreases, which means a deeper deformation, as can be seen in Fig. \ref{1dep}. Increasing the number of depressions is almost the same as increasing the number of bumps, except for the ends of the deformed tube as can be seen in Fig. \ref{ndep}. This has the effect of inverting the shifts observed in Fig. \ref{nbump}. As in the bump cases, one can note in Fig. \ref{1dep} $(c)$ and Fig. \ref{ndep} $(c)$  that the PDM  corrects the location of the resonant peaks of the transmittance and the width of the energy gap, respectively.  Furthermore, the inclusion of the PDM induces a shorter energy gap.

There is also another important result to be pointed out. In Ref. \cite{santos}, where the deformed nanotubes considered here were investigated without considering the PDM, was concluded that the pinched nanotubes induce a deeper geometrical potential compared with the nanotubes with bumps. However, as can be seen in our results, the inclusion of the PDM reveals that the opposite occurs. So, in order to induce a deeper geometrical potential, instead of creating a pinched nanotube, as was concluded in Ref. \cite{santos}, one should include bumps.

\section{Concluding Remarks}

We investigated the influence of the position dependent mass in the transport properties of curved two-dimensional quantum systems. Following the da Costa approach, we obtained the Schr\"odinger equation for quantum particles confined to a general surface with a PDM. We verified that the PDM introduces two extra terms in the Hamiltonian and that it does not modify the form of the geometrical potential, but changes its intensity. We found expressions for the transmittance and reflectance for a general surface of revolution, which can be obtained by solving a mixed set of differential and algebraic equations. As a first example of the general results obtained here, we investigate corrugated nanotubes. We verified that, even for a small deformation, as the number of corrugations increases the inclusion of the PDM becomes very relevant. In fact, the PDM corrects the location of the resonant peaks in the transmittance and the width of the energy gap. Also, we showed that neglecting the variation of the effective mass can lead to wrong conclusions about the transport properties in corrugated nanotubes. Then, we can conclude that the PDM plays an important role in the transport properties of two-dimensional quantum systems. Also, for future applications of curvature in the design of nanotube-based electronic devices, one might think of modeling nanotubes with desired specific properties with suitable choice of the function $m^{*}(y)$ via its shape.
 
\newpage
%
%\section*{Figures}

\newpage

\newpage

\newpage

\pagebreak[4]
% Non-BibTeX users please use
\bibliographystyle{elsarticle-num}
\bibliography{ref}

\end{document}